\documentclass[twocolumn,aps,showpacs,prb]{revtex4-1}
\usepackage{graphicx}
\usepackage{epstopdf}
\usepackage{amsmath}
\usepackage{multirow}

\begin{document}

\title{Perfect charge compensation in extremely large magnetoresistance materials LaSb and LaBi revealed by the first-principles calculations}

\author{Peng-Jie Guo$^{1,2}$}
\author{Huan-Cheng Yang$^{1,2}$}
\author{Kai Liu$^{1,2}$}\email{kliu@ruc.edu.cn}
\author{Zhong-Yi Lu$^{1,2}$}\email{zlu@ruc.edu.cn}

\affiliation{$^{1}$Department of Physics, Renmin University of China, Beijing 100872, China}
\affiliation{$^{2}$Beijing Key Laboratory of Opto-electronic Functional Materials $\&$ Micro-nano Devices, Renmin University of China, Beijing 100872, China}

\date{\today}

\begin{abstract}

By the first-principles electronic structure calculations, we have systematically studied the electronic structures of recently discovered extremely large magnetoresistance (XMR) materials LaSb and LaBi. We find that both LaSb and LaBi are semimetals with the electron and hole carriers in perfect balance. The calculated carrier densities in the order of $10^{20}$ cm$^{-3}$ are in good agreement with the experimental values, implying long mean free time of carriers and thus high carrier mobilities. With a semiclassical two-band model, the perfect charge compensation and high carrier mobilities naturally explain (i) the XMR observed in LaSb and LaBi; (ii) the non-saturating quadratic dependence of XMR on external magnetic field; and (iii) the resistivity plateau in the turn-on temperature behavior at very low temperatures. The explanation of these features without resorting to the topological effect indicates that they should be the common characteristics of all perfectly electron-hole compensated semimetals.

\end{abstract}


\maketitle

\section{INTRODUCTION}

The magnetoresistance (MR) effect, which describes the magnetic-field induced change of electrical resistance, not only continuously provides exciting physical phenomena since its discovery in 1857 \cite{Thomson57}, but also brings revolution to human beings' modern lives through its device applications such as hard drive \cite{daughton99} and magnetic sensor \cite{reig09}. Until now, many kinds of magnetoresistance effects have been identified. The conventional MR found in simple metals is of the order of a few percent \cite{pippard, stohr}. The other well-known MR effects, including the giant magnetoresistance (GMR) in magnetic multilayers \cite{fert88, grunberg89}, the colossal magnetoresistance (CMR) in perovskite manganites \cite{salamon01}, and the tunnel magnetoresistance (TMR) in magnetic tunnel junctions \cite{moodera95}, are more prominent than the conventional MR. Recently, much attention has been paid on the extremely large magnetoresistance (XMR) around 10$^4$ to 10$^6$ percent discovered in a few semimetals: WTe$_2$ \cite{ali14}, NbP \cite{shekhar15}, Cd$_3$As$_2$ \cite{liang15}, LaSb \cite{tafti15}, and etc. The most astonishing finding is that the XMR in WTe$_2$ doesn't saturate even under a magnetic field as high as 60 teslas \cite{ali14}.

Several mechanisms have been proposed to explain the XMR found in these semimetals. The quadratic dependence of MR on the perpendicular magnetic field without saturation in WTe$_2$ can be described by the perfect electron-hole compensation from a semiclassical two-band model \cite{ali14, pletikosic14}. The exotic longitudinal linear MR with parallel magnetic field and current directions in WTe$_2$ is explained from a quantum viewpoint by incorporating Landau levels \cite{wang15}. Previously, the quantum scenario \cite{abrikosov98} was also applied to the linear MR of nonstoichiometric silver chalcogenides in a perpendicular magnetic field \cite{xu97}. Besides, the suppression of backscattering channels at zero field and their opening under magnetic filed are suggested to play an important role in the magnetoresistance of WTe$_2$ \cite{feng15} and the Dirac semimetal Cd$_3$As$_2$ \cite{liang15}. Recently, similar to that found in WTe$_2$ \cite{ali14}, the non-saturating XMR with quadratic dependence on magnetic field were observed in two rare earth monopnictides LaSb \cite{tafti15, cava16} and LaBi \cite{sun16, kumar16, cava16}. Nevertheless, a consensus has not been achieved on the origin of their non-saturating XMR \cite{tafti15, sun16, kumar16, cava16}. To be specific, whether the semiclassical, quantum, or backscattering mechanism remains to be verified. Furthermore, LaSb and LaBi are predicted to be three-dimensional topological insulators \cite{zeng15}, which adds more concern to these lanthanum monopnictides.

In addition to sophisticated experimental tools, first-principles calculations have also played an important role in interpreting the electronic structures and the non-saturating XMR of WTe$_2$ \cite{ali14, feng15, sun15}. As to LaSb and LaBi, the previous calculations by using augmented plane wave (APW) method with two different formalisms, i.e., the local density approximation (LDA) and the Slater exchange potential, give distinct band structures for LaSb yet similar ones for LaBi \cite{hasegawa85}. The calculations based on the Slater exchange potential agree better with the de Haas-van Alphen effect measurements \cite{hasegawa85, kitazawa83,kitazawa82}. It has been known that for the exchange-correlation functionals, according to Perdew \textit{et al.}, there are several rungs in the 'Jacob's ladder' of density functional approximations \cite{perdew05}, which describe the materials' properties in different precision with increasing computating load. The LDA in the first rung with electron densities and the generalized gradient approximation (GGA) in the second rung with both electron densities and their gradients perform very well in studying the properties of a variety of systems \cite{perdew92}. Nevertheless, due to the derivative discontinuity of the exchange-correlation energy \cite{perdew83, sham83}, the LDA and GGA usually underestimate the band gaps of semiconductors. The meta-GGA in the third rung includes Laplacians of electron density and the kinetic energy densities beyond the GGA. It gives band gaps similar to the hybrid functional or GW methods, but with much less computational demand \cite{becke06, tran09}. The recently discovered XMR materials, including LaSb \cite{tafti15, cava16} and LaBi \cite{sun16, kumar16, cava16}, are semimetals with low carrier densities. Since their band gaps and carrier densities are between those of metals and semiconductors, whether the first-principles calculations at the GGA and the meta-GGA levels would give consistent conclusions on the XMR mechanism of semimetals LaSb and LaBi remains to be unraveled.

In this work, we report systematic studies on the electronic structures of LaSb and LaBi by using the first-principles electronic structure calculations. Two levels of exchange-correlation functionals, the GGA and the meta-GGA, are adopted respectively. Although they give somewhat different band structures and distinct band overlaps at certain $k$ point in Brillouin zone, the calculated Fermi surfaces and the derived ratio of electron and hole carrier densities point to the same conclusion that the extremely large MR of LaSb and LaBi originates from the perfect electron-hole compensation.

\section{COMPUTATIONAL DETAILS}

To study the electronic structures of LaSb and LaBi from first principles, we carried out calculations with the projector augmented wave (PAW) method \cite{paw1,paw2} as implemented in the VASP package \cite{vasp1,vasp2,vasp3}. For the exchange-correlation functional, we adopted two different levels in the Jacob's ladder \cite{perdew05}: the GGA and the meta-GGA. In the GGA level, the Perdew-Burke-Ernzerhof (PBE) formulae \cite{pbe} was used, while in the meta-GGA level, the modified Becke-Johnson (MBJ) exchange potential \cite{becke06, tran09} in combination with the GGA correlation was employed.

The kinetic energy cutoff of the plane wave basis was set to be 300 eV. For the Brillouin zone sampling, a 20$\times$20$\times$20 dense $k$-point mesh was utilized for the primitive cell, which contains one formula unit (f.u.), of these rock-salt structural crystals. The Gaussian smearing with a width of 0.05 eV was adopted around the Fermi surface. Both cell parameters and internal atomic positions were fully relaxed until all forces on atoms were smaller than 0.01 eV/\AA. The calculated lattice constants 6.56 \AA $ $ for LaSb and 6.67 \AA $ $ for LaBi agree quite well with the experimental values 6.49 \AA $ $ and 6.58 \AA $ $ \cite{leger84, sun15}, respectively. After the equilibrium structure was obtained, the electronic structures were calculated by including the spin-orbital-coupling (SOC) effect. To accurately describe the subtle electronic structures of LaSb and LaBi around the Fermi level, we have tested the 16$\times$16$\times$16, 20$\times$20$\times$20, and 24 $\times$24$\times$24 $k$-point meshes. We found the 20$\times$20$\times$20 $k$-point mesh renders converged band structures for the calculations based on the MBJ potential. The maximally localized Wannier functions (MLWF) \cite{wannier1, wannier2} were then used to calculate the Fermi surfaces and the carrier concentrations were analyzed based on the information of Fermi surface volumes.

\section{RESULTS AND ANALYSIS}

\subsection{LaSb}

\begin{figure}[tbp]
\includegraphics[angle=0,scale=0.4]{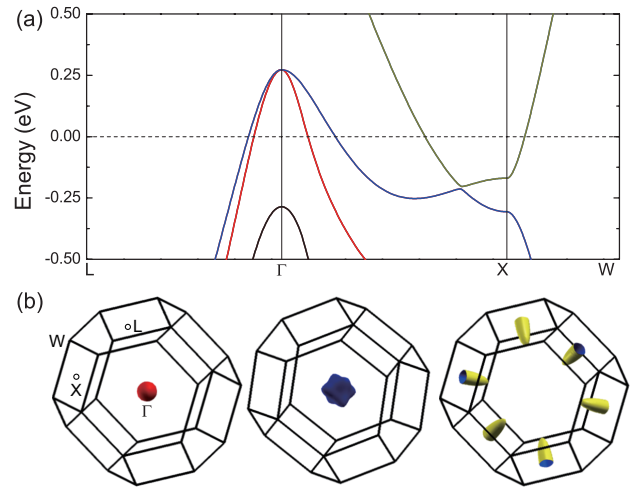}
\caption{(Color online) (a) Band structures along high symmetry directions of Brillouin zone and (b) Fermi surfaces of LaSb calculated with the PBE functional and the SOC effect.}
\label{fig1}
\end{figure}

Figure 1 shows the band structure along high symmetry directions of Brillouin zone (BZ) and the Fermi surface of LaSb calculated with the PBE functional and including the SOC effect. Here we only focus on those bands around the Fermi level. We have checked that except for those bands displayed in Fig. 1(a), there is no band crossing the Fermi level along other high symmetry directions in BZ, demonstrating the semimetallic behavior of LaSb. As can be seen, there are two doubly-degenerate bands across the Fermi level around the $\Gamma$ point and one doubly-degenerate band around the $X$ point. Inclusion of the SOC effect opens up a tiny gap at the anti-crossing point of bands along the $\Gamma$-$X$ direction. This is in accordance with the previous band structure calculations on LaSb \cite{zeng15}. The corresponding Fermi surface sheets of these bands, whose colors are in one-to-one relationship, are given in Fig. 1(b). There are two hole pockets around the $\Gamma$ point: the smaller one takes an isotropic spherical shape; the bigger one looks like three crossing spindles. The electron pockets are around the $X$ points of the first BZ. When we move half of these pockets to the equivalent opposite $X$ points, they will form three ellipsoids with the long axis pointing along the $\Gamma$-$X$ direction. By computing the volumes of the electron and hole pockets, we obtain the hole-type carrier densities as 2.10$\times$10$^{20}$ cm$^{-3}$ and the electron-type carrier densities as 2.20$\times$10$^{20}$ cm$^{-3}$ (Table I). This is in the same order of magnitude as that measured in transport experiment \cite{tafti15}. The ratio between the densities of electron-type carriers and hole-type carriers is 1.05, indicating a perfect compensation between the electrons and the holes.

\begin{figure}[tbp]
\includegraphics[angle=0,scale=0.4]{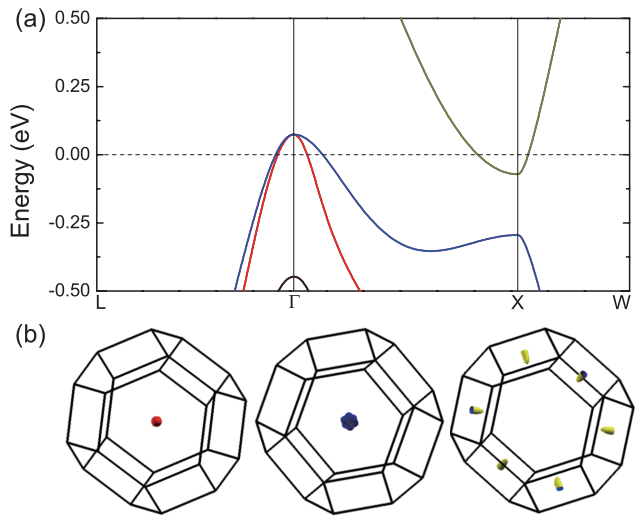}
\caption{(Color online) (a) Band structures along high symmetry directions of Brillouin zone and (b) Fermi surfaces of LaSb calculated with the MBJ potential and the SOC effect.}
\label{fig2}
\end{figure}

\begin{table}[b]
\caption{Electron-type ($\alpha$ band) and hole-type ($\beta$ and $\gamma$ bands) carrier densities (in unit of 10$^{20}$ cm$^{-3}$) and their ratios for LaSb and LaBi calculated with the PBE functional at the GGA level and the MBJ potential at the meta-GGA level.}
\begin{center}
\begin{tabular*}{8cm}{@{\extracolsep{\fill}} ccccc}
\hline \hline
 & \multicolumn{2}{c}{LaSb} & \multicolumn{2}{c}{LaBi} \\
\cline{2-3} \cline{4-5}
 & PBE & MBJ & PBE & MBJ \\
\hline
$n_e$($\alpha$) & 2.20 & 0.38 & 3.45 & 1.65 \\
$n_h$($\beta$) & 0.43 & 0.08 & 0.71 & 0.31 \\
$n_h$($\gamma$) & 1.67 & 0.32 & 2.95 & 1.31 \\
$N_e/N_h$ & 1.05 & 0.95 & 0.94 & 1.02 \\
\hline \hline
\end{tabular*}
\end{center}
\end{table}

\begin{figure}[t]
\includegraphics[angle=0,scale=0.4]{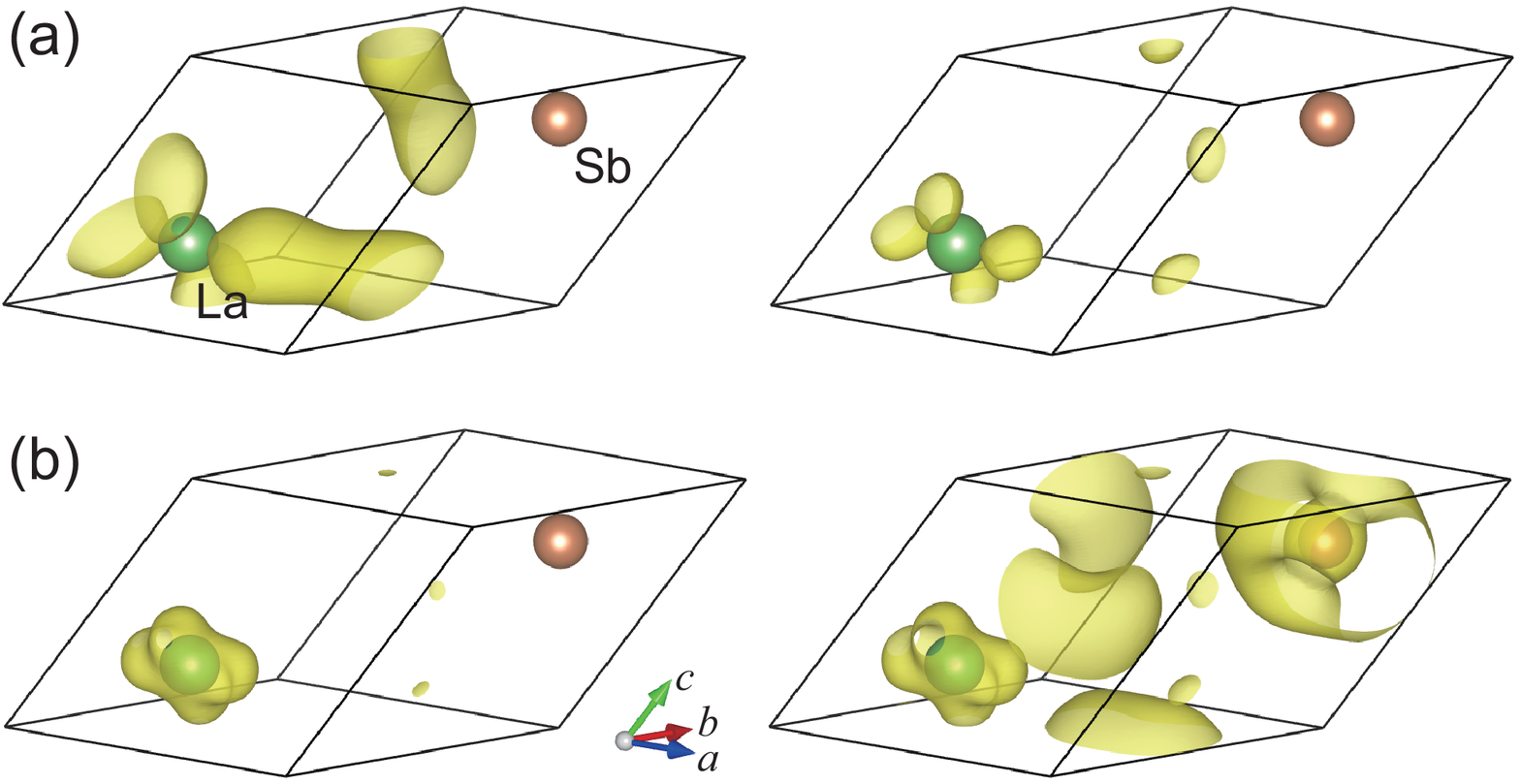}
\caption{(Color online) Band decomposed charge densities of LaSb for two doubly-degenerate bands with energies (a) -0.31 eV and (b) -0.17 eV below Fermi level at the $X$ point of Fig. 1(a) calculated with the PBE functional and the SOC effect.}
\label{fig3}
\end{figure}

\begin{figure}[b]
\includegraphics[angle=0,scale=0.4]{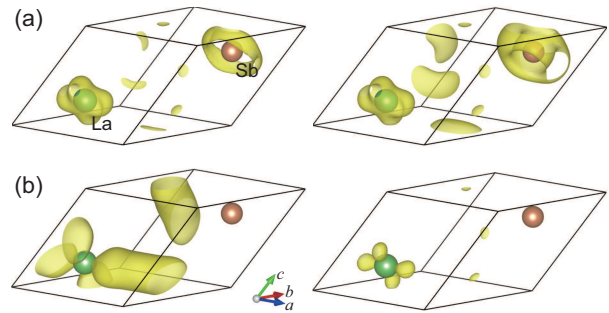}
\caption{(Color online) Band decomposed charge densities of LaSb for two doubly-degenerate bands with energies (a) -0.29 eV and (b) -0.07 eV below Fermi level at the $X$ point of Fig. 2(a) calculated with the MBJ potential and the SOC effect.}
\label{fig4}
\end{figure}

As LaSb is a semimetal, we have also studied its band structure and Fermi surface with the MBJ potential in the meta-GGA level. Compared with the band structure calculated with the PBE functional [Fig. 1(a)], dramatic changes in Fig. 2(a) can be discerned. The top of two doubly-degenerate valence bands at $\Gamma$ point shift downwards about 0.2 eV, while the bottom of the doubly-degenerate conduction band at the $X$ point shifts upwards. These band shifts yield obvious reduction of the Fermi surface volumes [Fig. 2(b)] and thus the carrier densities (Table I). Moreover, in the MBJ calculations, there is no anti-crossing of bands along the $\Gamma$-$X$ direction, which seems to lift the overlap between valence band and conduction band in the PBE calculations [Fig. 1(a)]. Similar phenomenon was also noticed by Tafti \textit{et al.} \cite{tafti15}. More information on the bands below the Fermi level around the $X$ point can be obtained from their corresponding charge densities as in the following. On the other hand, although the carrier densities reduce sharply, the elaborate calculation using dense $k$ points in the whole BZ renders the carrier-densities ratio between electrons and holes as 0.95, which is in good accordance with the ratio [0.014/(0.004+0.011)=0.93] measured in previous the de Hass-van Alphen experiment \cite{hasegawa85,kitazawa82}. Thus no matter whether the calculations are using the PBE functional at the GGA level or the MBJ potential at the meta-GGA level (Table I), the perfect charge compensation in LaSb always holds.

To analyze the changes of band characteristics at the $X$ point of Brillouin zone (Figs. 1 and 2), we have plotted the band decomposed charge densities calculated with the PBE functional in Figure 3 and calculated with the MBJ potential in Figure 4, respectively. For the two doubly-degenerate bands at the $X$ point shown in Fig. 1(a), the charge densities for the bands with energy -0.31 eV below Fermi level mainly distribute around the La atom [shown in Fig. 3(a)]. On the other hand, the bands with energy -0.17 eV below Fermi level at $X$ point [Fig. 1(a)] demonstrate some charge distributions around the Sb atom [Fig. 3(b)]. The higher-energy (-0.17 eV) conduction band with charges on the anion atom and the lower-energy (-0.31 eV) valence band with charges on the cation atom imply a band inversion at the $X$ point, as also revealed by Zeng \textit{et al.} \cite{zeng15}. However, when the MBJ potential is adopted, the above band characteristics change thoroughly. For the two doubly-degenerate bands with energies -0.29 eV and -0.07 eV at the $X$ point [shown in Fig. 2(a)], their corresponding charge densities are displayed in Figs. 4(a) and 4(b), respectively. The lower-energy (-0.29 eV) valence band with charges on the anion atom Sb [Fig. 4(a)] and the higher-energy (-0.07 eV) conduction band with charges on the cation atom La [Fig. 4(b)] indicate no band inversion at the $X$ point. Whether the calculations with the PBE functional or those with the MBJ potential offers the correct band structures of LaSb around the $X$ point needs experimental verification such as angle-resolved photoemission spectroscopy (ARPES) measurement.

\subsection{LaBi}

\begin{figure}[tbp]
\includegraphics[angle=0,scale=0.4]{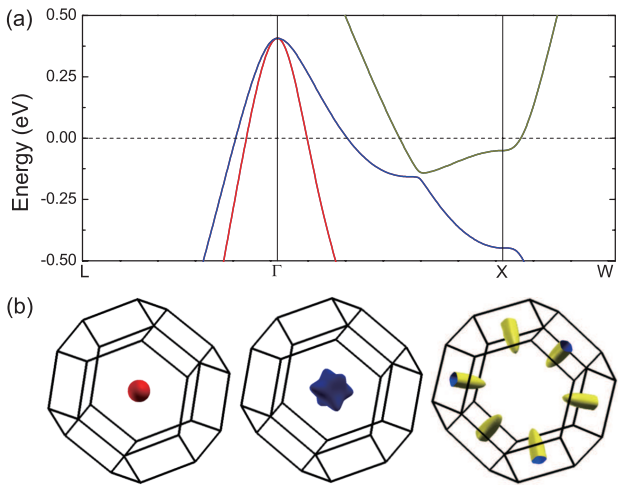}
\caption{(Color online) (a) Band structures along high symmetry directions of Brillouin zone and (b) Fermi surfaces of LaBi calculated with the PBE functional and the SOC effect.}
\label{fig5}
\end{figure}

The band structure and Fermi surface of LaBi calculated with the PBE functional and including the SOC effect are shown in Figure 5. In the element table, Bi belongs to the same main group as Sb, but with a heavier mass and a larger atomic radius. Compared with the 5$p$ orbitals of Sb, the Bi 6$p$ orbitals are more extended. This will induce more wavefunction overlap with the La atom when forming the rock-salt structural crystal LaBi, as reflected in the higher energies of the valence band top around $\Gamma$ point [Fig. 5(a)] and the larger overlap area between two anti-crossing bands around the $X$ point than that of LaSb [Fig. 1(a)]. In addition, due to the heavier mass of Bi than Sb, the SOC effect in LaBi is more prominent. As a result, the band gap along the $\Gamma$-$X$ direction is more notable [Fig. 5(a)]. These features in the band structure of LaBi are consistent with those reported in the previous calculations using the GGA functional \cite{zeng15}. Meanwhile, the Fermi surface volume of LaBi [Fig. 5(b)] is larger than that of LaSb [Fig. 1(b)], which leads to higher carrier densities ($\sim$3.5$\times$10$^{20}$ cm$^{-3}$) for both electrons and holes (Table I). The calculated densities are in the same order of magnitude with the values estimated from magnetotransport measurements \cite{sun16}. The ratio between the densities of electron-type and hole-type carriers calculated with the PBE functional is 0.94, revealing a perfect electron-hole balance in LaBi as well \cite{sun16}.

\begin{figure}[tbp]
\includegraphics[angle=0,scale=0.4]{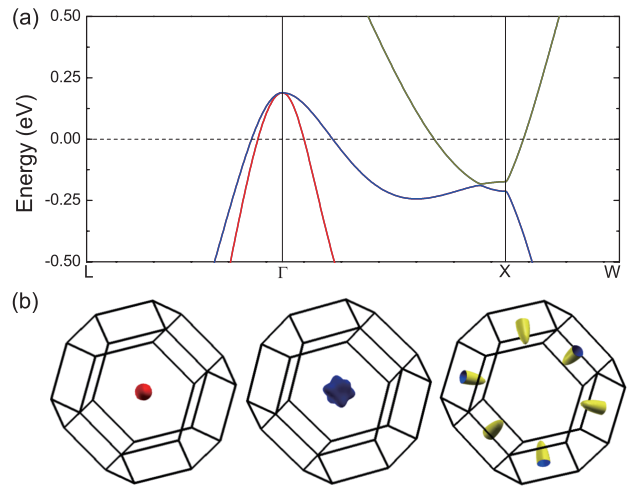}
\caption{(Color online) (a) Band structures along high symmetry directions of Brillouin zone and (b) Fermi surfaces of LaBi calculated with the MBJ potential and the SOC effect.}
\label{fig6}
\end{figure}

Similar to LaSb, when the MBJ potential is applied in the calculations, the band structure and Fermi surface of LaBi also show some obvious changes (Fig. 6). Compared with the PBE results [Fig. 5(a)], downward shifting for the top of valence bands at $\Gamma$ point and upward shifting for the bottom of conduction band at the $X$ point are found. However, due to the larger overlap between the valence and conduction bands in LaBi than that of LaSb, the introduction of the MBJ potential does not eliminate the overlap completely, as indicated by the reserved band overlap below the Fermi level at the $X$ point [Fig. 6(a)]. On the other hand, the band shifts cause the reduction of both Fermi surface volume [Fig. 6(b)] and carrier densities (Table I). Nevertheless, the calculated density ratio between electrons and holes is 1.02 (Table I), i.e., in a perfect charge compensation as well. This is in accordance with the measured ratio [0.027/(0.0064+0.022)=0.95] in the previous de Hass-van Alphen experiment \cite{hasegawa85,kitazawa82}.

\begin{figure}[t]
\includegraphics[angle=0,scale=0.4]{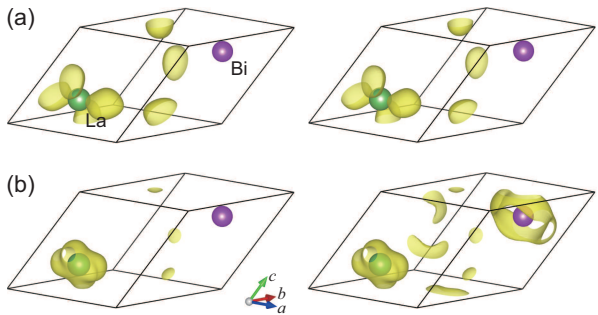}
\caption{(Color online) Band decomposed charge densities of LaBi for two doubly-degenerate bands with energies (a) -0.21 eV and (b) -0.17 eV below Fermi level at the $X$ point of Fig. 6(a) calculated with the MBJ potential and the SOC effect.}
\label{fig7}
\end{figure}

Since both the PBE functional and the MBJ potential calculations reserve the band overlap around the $X$ point of BZ for LaBi (Figs. 5 and 6), we plot in Figure 7 the band decomposed charge densities calculated with the MBJ potential for illustration. The lower-energy (-0.21 eV) band displays charge distributions around the cation atom La [Fig. 7(a)], while the higher-energy (-0.17 eV) band also shows some charge distributions around the anion atom Bi [Fig. 7(b)]. Thus for LaBi, a band inversion around the $X$ point exists, which agrees with the findings by Zeng \textit{et al.} \cite{zeng15} and Hasegawa \cite{hasegawa85}. In addition to the perfect charge compensation (Table I), the topological effect in LaBi may introduce some interesting topological surface state on certain surface \cite{zeng15}, which may interact with its XMR anisotropically.

\section{DISCUSSION}

From the semiclassical two-band model, the longitudinal electrical resistivity of nonmagnetic materials with both electron and hole carriers under magnetic field reads \cite{mermin,ziman}:
\begin{equation}
\rho(B)=\frac{(n_e\mu_e+n_h\mu_h)+(n_e\mu_h+n_h\mu_e)\mu_e\mu_hB^2}{e(n_e\mu_e+n_h\mu_h)^2+e(\mu_e\mu_h)^2(n_e-n_h)^2B^2}
\end{equation}
where $n_e$ ($n_h$) is the electron (hole) concentration, $\mu_e$ ($\mu_h$) the mobility of electrons (holes), $e$ the charge unit of electron, and $B$ the magnetic field. Then the magnetoresistance, which describes the change of electrical resistance in response to the magnetic field, yields
\begin{eqnarray}
MR&=&\frac{\rho(B)-\rho(0)}{\rho(0)} \nonumber \\
&=&\frac{n_e\mu_en_h\mu_h(\mu_e+\mu_h)^2B^2}{(n_e\mu_e+n_h\mu_h)^2+(n_e-n_h)^2(\mu_e\mu_h)^2B^2}
\end{eqnarray}
When the perfect charge compensation is satisfied ($n_e$=$n_h$), the MR shows a quadratic dependence on the magnetic field with the product of electron and hole mobilities as coefficient:
\begin{equation}
MR=\mu_e\mu_hB^2
\end{equation}

We first concentrate on the magnetoresistances of LaSb and LaBi as well as their relationship with the magnetic field. Our above calculations have shown that the electron and hole carriers fulfill the perfect compensation (Table I), thus Eq. (3) can be applied. This naturally explains the quadratic dependence of MR on magnetic field and its non-saturating behavior observed in LaSb \cite{tafti15, cava16} and LaBi \cite{sun16, kumar16, cava16}. On the other hand, Eq. (3) indicates that the MR is proportional to the carrier mobility product. Since the mobility $\mu$ equals to $e\tau/m^*$, the longer is the mean free time $\tau$ and the smaller is the carrier effective mass $m^*$, the higher the mobility $\mu$ would be. Our calculated carrier densities of semimetals LaSb and LaBi are quite low (Table I), thus the carriers would have a long mean free time and thus high mobility at low temperatures. In fact, the derived mobilities of LaSb and LaBi from low-temperature transport experiments do manifest high values ($10^4 \sim 10^5$ cm$^2$V$^{-1}$s$^{-1}$) \cite{tafti15, sun16, kumar16} comparable to that found in WTe$_2$ \cite{ali14,tafti15}. From Eq. (3), these high carrier mobilities would result in extremely large MR, which are observed in LaSb \cite{tafti15,cava16} and LaBi \cite{sun16, kumar16, cava16}. Our calculated results with perfect charge compensation and low carrier densities (Table I) consistently interpret the XMR and its non-saturating quadratic magnetic-field dependence found in LaSb and LaBi.

The temperature effect is prominent in the recently found nonmagnetic XMR materials, which all demonstrate metallic behavior at high temperatures and turn-on of MR at low temperatures \cite{ali14, shekhar15,liang15,tafti15}. The origin of the resistivity enhancement at low temperatures has been well addressed by Wang \textit{et al.}: instead of a magnetic-field-induced metal-insulator transition, it is due to the opposite temperature dependences of the ordinary resistivity at zero magnetic field and the other part induced by magnetic field \cite{wangyl15}. Another interesting phenomenon found in lanthanum monopnictides is the resistivity plateau below $\sim$15 K \cite{tafti15, sun16}. Previously, the resistivity plateau in LaSb was compared with a similar plateau protected by the conducting surface states of a topological insulator candidate SmB$_6$ \cite{tafti15}. In fact, the resistivity plateau in LaSb can also be understood within the framework of semiclassical two-band model as in LaBi \cite{sun16}. Our calculations show that LaSb is in perfect electron-hole balance $n_e=n_h=n$ (Table I), then from Eq. (1), the resistivity becomes:
\begin{equation}
\rho(T,B)=\frac{1+\mu_e(T)\mu_h(T)B^2}{en(T)(\mu_e(T)+\mu_h(T))}
\end{equation}
The dependence of resistivity $\rho(T,B)$ on temperature $T$ is included in the temperature-dependent carrier densities $n(T)$ and mobilities $\mu(T)$. At very low temperatures without thermal excitation, the carrier densities are nearly constant. On the other hand, as mentioned above, the mobility equals to $e\tau/m^*$. Below very low temperatures: the lattice parameters and band structures change little, then the effective mass $m^*$ of carriers will not show much variation; moreover, the crystal lattice is frozen and the mean free time $\tau$ depends on carrier scattering. The $\tau$ is inversely proportional to the carrier densities, which will approach to a constant at very low temperatures. Actually, the derived carrier densities and mobilities of LaBi from transport measurements indeed access to constants at very low temperatures \cite{sun16}. As a result, the nearly constant $n$ and $\mu$ will induce a resistivity plateau under very low temperature, which is observed in LaSb \cite{tafti15}. 

The carrier densities in semimetals depend sensitively on the Fermi level. Due to the low carrier densities of LaSb and LaBi (Table I), a slight change in Fermi level would induce large variation of the density ratio between different types of carriers. We take the LaSb for example. Within the PBE functional calculations, a 10 meV (corresponding to a temperature of 116 K) downward shift of Fermi level would change the ratio $N_e$/$N_h$ from 1.05 to 0.79, while a 20 meV shift yields 0.71. With the MBJ potential, a 10 meV downward shift of Fermi level would vary the ratio from 0.95 to 0.59 and a 20 meV shift to 0.33. This obvious change of charge compensation would have a substantial reflection in the MR. The susceptible carrier ratio in LaSb to Fermi level draws our attention when interpreting the experimental findings, since surface adsorption or surface vacancies after exfoliation can easily tune the potential for the surfacial layers of LaSb and LaBi. Instead, compared with the three-dimensional rock-salt structures of LaSb and LaBi, the layered structure of bulk WTe$_2$ with weak interlayer interaction makes its interior more inert to surface adsorption. In other words, the non-saturating quadratic XMR of bulk WTe$_2$ is more robust.

One may notice that until now, we have not resorted to the topological effect when elucidating the XMR found in LaSb and LaBi. In our calculations using the MBJ potential, LaSb does not show band inversion around the $X$ point (Figs. 2 and 4) while LaBi does (Figs. 6 and 7), thus these two lanthanum monopnictides may own different topological properties. However, due to their perfect charge compensation, low carrier densities (Table I), and high carrier mobilities \cite{tafti15, sun16, kumar16}, they both demonstrate non-saturating quadratic XMR and resistivity plateau at very low temperatures \cite{tafti15, sun16, kumar16}. Since these magnetic-field- and temperature-dependent behaviors can be well understood within the semiclassical two-band model, they should be the common features of perfectly electron-hole compensated semimetals, no matter whether the materials are topological or not. 

\section{CONCLUSION}

By using first-principles calculations, we have systematically studied the electronic structures of the recently discovered XMR semimetals LaSb and LaBi. In spite of two different levels of exchange-correlation energy, the PBE functional in the GGA level and the MBJ potential in the meta-GGA level, are adopted, we draw the same conclusions: (i) both LaSb and LaBi are in perfect electron-hole compensation; (ii) the carrier densities are in the order of $10^{20}$ cm$^{-3}$ as semimetals. From the semiclassical two-band model, these features combined with the high carrier mobilities derived from transport experiments \cite{tafti15,sun16,kumar16} naturally explain the non-saturating quadratic XMR found in LaSb and LaBi as well as the resistivity plateau at very low temperatures. We also find that the charge compensation in the semimetals LaSb and LaBi depends sensitively on the Fermi level position, thus one must be careful when interpreting related experimental data.

On the other hand, the PBE calculations on LaSb give inverse bands at the $X$ point of Brillouin zone while the MBJ calculations show no inversion. In comparison, due to a larger overlap between the valence and conduction bands in LaBi, the usage of different types of exchange-correlation functional yields the same band inversion around the $X$ point. Thus LaSb and LaBi may possess different topological properties. The correct band features of LaSb and LaBi around the $X$ point need to be verified by experiments such as angle-resolved photoemission spectroscopy (ARPES) measurement.

Our studies suggest that in spite of the topological properties, the non-saturating quadratic XMR and the resistivity plateau at very low temperatures should be the universal behaviors of all semimetals in perfect electron-hole balance, for which these features can be captured by the semiclassical two-band model. Our first-principles calculations revealing the perfect charge compensation in both LaSb and LaBi are very crucial to accurately understand their XMR phenonmena found in experiments.

\begin{acknowledgments}

We thank Hechang Lei, Shancai Wang, and Tian Qian for helpful discussions. This work was supported by the National Natural Science Foundation of China (Grants 11190024 and 91421304). KL was supported by the Fundamental Research Funds for the Central Universities, and the Research Funds of Renmin University of China (14XNLQ03). Computational resources have been provided by the Physical Laboratory of High Performance Computing at RUC. The Fermi surfaces were prepared with the XCRYSDEN program \cite{kokalj03}.

\end{acknowledgments}

\end{document}